\documentclass[preprint,12pt]{elsarticle}

\usepackage{graphicx}
\usepackage{amssymb}

\usepackage{placeins}
\usepackage{url}
\usepackage[hidelinks, colorlinks=false]{hyperref}
\usepackage{amsmath}
\usepackage{tikz}
\usetikzlibrary{shapes.geometric, arrows}
\tikzstyle{startstop} = [rectangle, minimum width=1cm, minimum height=1cm,text centered, draw=black,]
\tikzstyle{io} = [trapezium, trapezium left angle=70, trapezium right angle=110, minimum width=3cm, minimum height=1cm, text centered, draw=black, fill=blue!30]
\tikzstyle{process} = [rectangle, minimum width=3cm, minimum height=1cm, text centered, draw=black, fill=orange!30]
\tikzstyle{decision} = [diamond, minimum width=3cm, minimum height=1cm, text centered, draw=black, fill=green!30]
\tikzstyle{arrow} = [thick,->,>=stealth]

\journal{Physica A}

\begin{document}

\begin{frontmatter}


\title{Modeling the Infectiousness of Twitter Hashtags}



\author[Skaza]{Jonathan Skaza\corref{cor1}}
\author[Blais1,Blais2]{Brian Blais}

\address[Skaza]{Department of Biostatistics, University of Michigan, United States}
\address[Blais1]{Department of Science and Technology, Bryant University, United States}
\address[Blais2]{Institute for Brain and Neural Systems, Brown University, United States}
\cortext[cor1]{Corresponding author at {\tt jskaza@umich.edu}}

\begin{abstract}
This study applies dynamical and statistical modeling techniques to quantify the proliferation and popularity of trending hashtags on Twitter.
Using time-series data reflecting actual tweets in New York City and San Francisco, we present estimates for the dynamics (i.e., rates of infection and recovery) of several hundred trending hashtags using an epidemic modeling framework coupled with Bayesian Markov Chain Monte Carlo (MCMC) methods. 
This methodological strategy is an extension of techniques traditionally used to model the spread of infectious disease.  We demonstrate that in some models, hashtags can be grouped by infectiousness, possibly providing a method for quantifying the trendiness of a topic.
\end{abstract}

\begin{keyword}
Twitter dynamics \sep trending \sep SIR \sep SIRI \sep MCMC \sep information diffusion


\end{keyword}

\end{frontmatter}



\section{Introduction}
\label{S:1}

Twitter  (\texttt{http://twitter.com}) is a popular social networking website that allows users to both send and read messages known as tweets.
The social networking site has approximately 320 million monthly active users that produce an average of about 500 million tweets per day \cite{about}.
Twitter serves as a place for users to share anything and everything on their minds---news stories, ideas, quotes, lyrics, etc.
Users are able to embed hashtags within their tweets by using the hash character (i.e., \texttt{\#}). 
Hashtags are metadata tags which allow tweets containing the text following the hash character to be grouped together. 
From there, it is possible for users to query certain hashtags to see what is being discussed throughout the site. 
The Twitter site even contains a panel of trending topics---hashtags and topics that have become very popular in a short period of time.

Hashtags can also prove useful for researchers in need of categorizing or grouping tweets.
While it is also possible to filter tweets by words or phrases, such an approach can be problematic.
For instance, a researcher interested in exploring the degree of happiness on Twitter may search for tweets containing the word ``happy''.
While this strategy will return tweets from people expressing sentiments such as ``I am {\it happy}'', it will also return messages in the category of ``I am not {\it happy}''.
Sentiment analysis techniques are needed to rectify this issue \cite{agarwal11}.
Using hashtags to filter tweets hedges against the need to address such concerns; a person who inserts \texttt{\#happy} into his/her tweet is likely happy.
However, the volume of tweets meeting the specific search criteria will be reduced because not every ``happy'' tweet, for example, will include \texttt{\#happy}.
Nevertheless, to avoid problems with contradicting sentiments, the present study uses hashtags as a proxy to study the prevalence and popularity of topics on Twitter.

  
This study attempts to quantify the spread of certain trending hashtags on Twitter. Using the methods described below, one can estimate the rates of infection and recovery for a particular trending topic.
Furthermore, with slight data processing, the same methodology can be used in a predictive context.
This study provides a brief overview of the existing literature concerning epidemic modeling and its use describing information dynamics on Twitter.
Subsequently, we describe the methods that we used to quantify the propagation of trending topics on Twitter.  In the process we find two main categories of hashtag dynamics---marginally infectious and very infectious, without much in between.

\subsection{Previous Work}
\label{S:1.1}

\begin{figure}[t]
\centering
\begin{tabular}{lr}
{\Large (A)}&\\
&\includegraphics[width=2in]{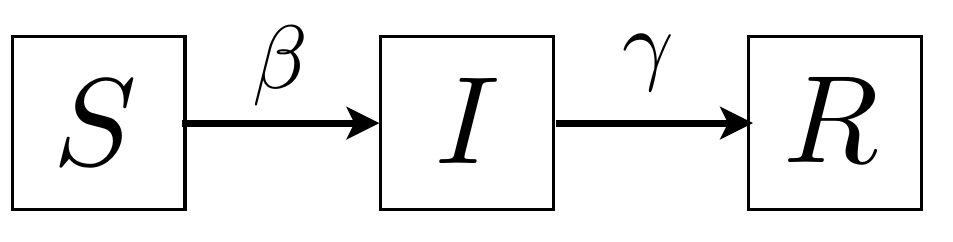}\\
{\Large (B)}&\\
&\includegraphics[width=2in]{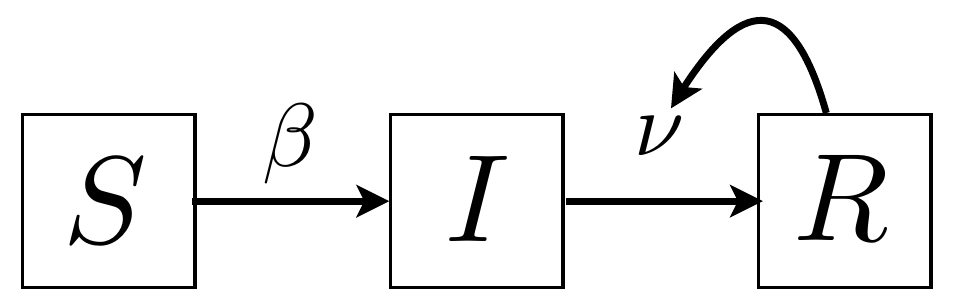}
\end{tabular}
\caption{Infection compartmental models.  Shown are the simple SIR model (A) and the more complex SIRI model (B).  }
\label{fig:sir_mod}
\end{figure}

Mathematical models have been used in the prediction, control, and analysis of epidemic phenomena---most notably, the spread of infectious disease throughout a population---since the advent of the susceptible, infected, and recovered (SIR) model \cite{kermack27}.
These types of epidemic models are featured in studies concerning measles \cite{mcgilchrist96,grais06,tuckwell07,kuniya06} and influenza \cite{tuckwell07,li09,hooten10,coelho11}, among others.
The basic SIR model describes the dynamical process of disease by categorizing members of the population of interest as either susceptible (S), infected (I), or recovered (R), while incorporating rates of infectiousness ($\beta$) and recovery ($\gamma$). 
Figure~\ref{fig:sir_mod}A illustrates a model diagram. 

Members of the population transition to and from different compartments based on the system of differential equations presented in Equation~\ref{eq:sir_diff}.

\begin{align} \label{eq:sir_diff}
\begin{split}
\frac{dS}{dt} =& -{\beta SI/N}\\
\frac{dI}{dt} =& +{\beta SI/N} - \gamma I\\
\frac{dR}{dt} =& +\gamma I
\end{split}
\end{align}

Although possibly useful for describing the dynamics of Twitter, it may be that the \textit{infection} of ideas doesn't follow the same structure.  Perhaps, as in \cite{DBLP:journals/corr/CannarellaS14}, the ``recovery'' from infection isn't a passive time-decay but depends on how many have recovered already.  In that case, referred to here as the SIRI model, we have a slightly modified set of equations which can lead to a more rapid decrease in the recovery phase as evidenced in Equation~\ref{eq:siri_diff}.

\begin{align} \label{eq:siri_diff} 
\begin{split}
\frac{dS}{dt} =& -\beta I S /N \\
\frac{dI}{dt} =& +\beta I S /N -\nu I R/N\\
\frac{dR}{dt} =& +\nu I R
\end{split}
\end{align}


A relatively new strategy in the field of epidemic modeling is to develop a statistical model for the parameters of the dynamical model.
Specifically, we use Markov Chain Monte Carlo (MCMC) simulation to estimate the posterior probabilities of the epidemic model's parameters (e.g., $\beta$, $\gamma$, and $\nu$) \cite{coelho11,blais13}.  

As mentioned above, much of the previous work featuring epidemic modeling techniques concerns the transmission of an infectious disease.
However, epidemic models may just as well be applied to capture the transmission of some other trait from individual to individual or group to group. 
A trait may take the form of a genetic characteristic, a cultural phenomenon, an addictive activity, the gain or loss of information, etc. \cite{brauer01}.

In this regard, several have applied SIR-type techniques to the spread of information on Twitter and other networks \cite{bettencourt06, jin13, zhao12, eager15}.
It is important to note that the present study concerns the dynamics of hashtag \emph{popularity} on Twitter.
We define a hashtag's popularity by how many users are tweeting about a topic at a given time.
Modeling popularity is different from other Twitter studies that address network phenomena \cite{bakshy11,lerman10,weng12,weng13,wu11} or specific behaviors such as information-sharing \cite{ko2014}.  The use of popularity as a measure greatly simplifies both the data-taking procedure and the  analysis, while at the same time providing a quantitative framework for discussing the infectiousness of ideas.

\section{Methods}
\label{S:2}

To apply the SIR model to real data requires a dataset that measures hashtag prevalence throughout Twitter over time. 
Therefore, the first stage of the present study's analysis of Twitter consists of developing a database of hashtags versus time.
We develop such a database using one of Twitter's application programming interfaces (APIs) and the Python programming language.  Twitter has a number of different APIs; however, the class of streaming APIs is most applicable to this study, as tweets are collected in near real-time \cite{apidoc}. 

Hence, by accessing the streaming API and performing a filter, we collect a sample of time-stamped tweets, each of which contains the hash character.  
The streaming API is also used to collect a geo-coded trending topics list.
This is done for New York City and San Francisco, for convenience.
Given a sample of time-stamped tweets, in the form of single stochastic events, the time series is then smoothed using a time-windowed average.  
Several time-windows are explored, and the results follow, but the majority of the averaged time-series perform best with an average of one hour.  
We have observed that this is the time-scale over which most Twitter activity happens.

\begin{figure} \label{fig:infectious-occurrence} 
\centering
\includegraphics[width=\linewidth]{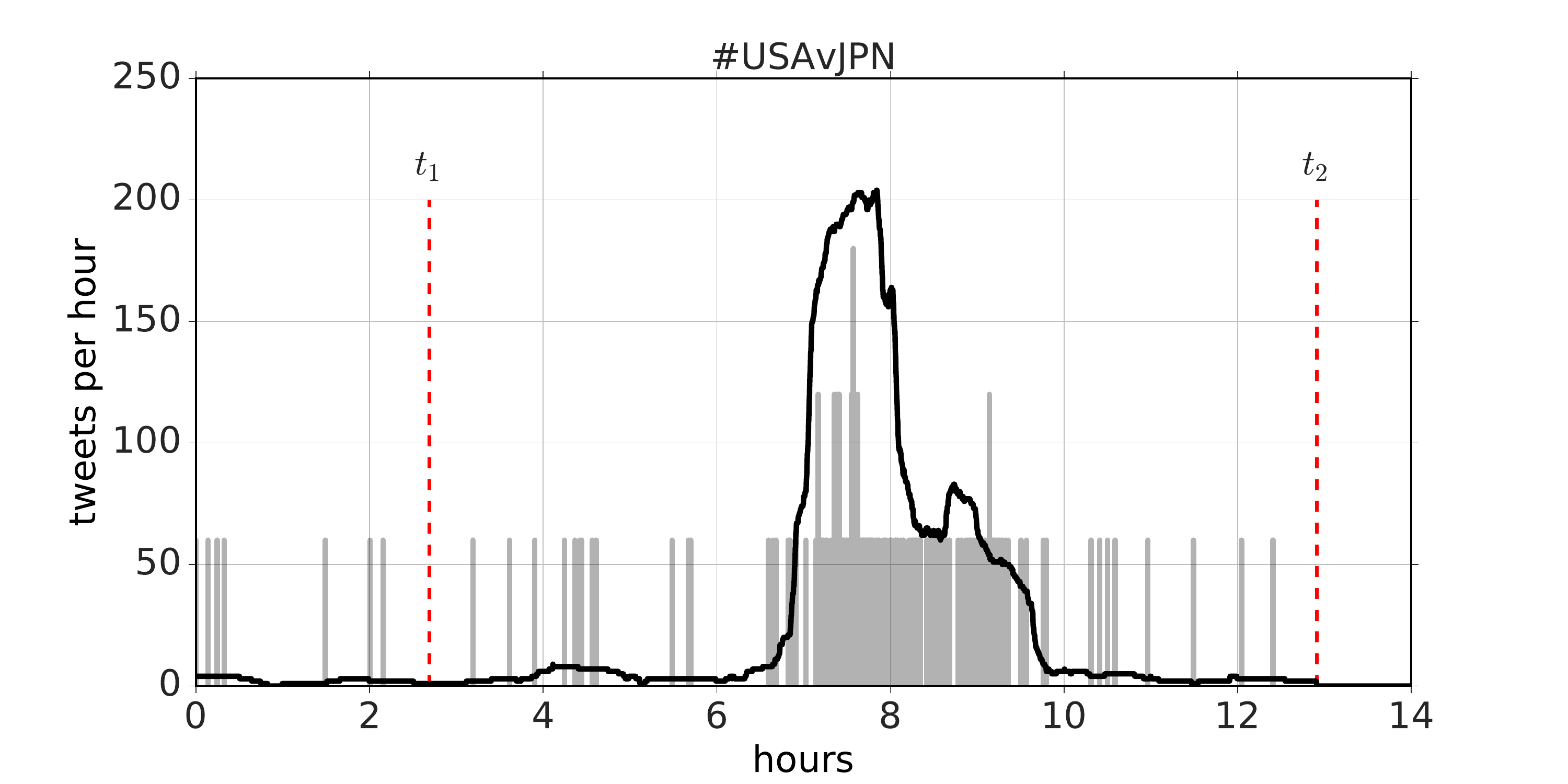}
\caption{Sample discrete tweet time series (gray) to smoothed continuous time series (black) for a hashtag, \texttt{\#USAvJPN}.  The automatically extracted subseries is between times $t_1$ and $t_2$ shown.  The smoothed time-scale is 1 hour.}
\end{figure}

Although a few series are selected by hand for examples, to automate the process we automatically scan the date and extract \textit{peaks} in activity, which represent an infectious occurrence.  
One such occurrence is defined in a time series with the following procedure:
\begin{enumerate}
\item find the maximum point in the time series
\item from that maximum, moving \textit{backwards} in time, find the point where the series is 1/100 of the maximum - call this time $t_1$
\item from that maximum, moving \textit{forwards} in time, find the point where the series is 1/100 of the maximum - call this time $t_2$
\end{enumerate}
The infectious occurrence is defined to be the time-series between  $t_1$ and  $t_2$, as shown in Figure~\ref{fig:infectious-occurrence}.

For each of these time-series, the infection model analysis is performed.   We describe this analysis presently. 

\subsection{Dynamical Model}
\label{sec:dynmodel}

In the present study, each dynamical model presented is either of the SIR (Equation~\ref{eq:sir_diff}) or SIRI (Equation~\ref{eq:siri_diff}) type, as described above.  Other forms could be explored but we feel they would not lead to anything more informative than these. The SIR and SIRI simulations were implemented using a wrapper, called \textit{pyndamics}, around the \texttt{odeint} function in Python's SciPy library \cite{scipy}.

\subsection{Reproduction Number}

The reproduction number for the SIR model, traditionally and confusingly denoted by $R$, is a threshold number related to the dynamical parameters of the SIR model above which the infection will grow.  We will refer to this number as ${\cal R}$ so that it is not confused with the recovered population, $R$.  It is easily derived from the dynamical equations,
\begin{align}
\begin{split}
\frac{dS}{dt} &= -\beta I S /N \\
\frac{dI}{dt} &= +\beta I S /N -\gamma I \\
\frac{dR}{dt} &= +\gamma I
\end{split}
\end{align}
The condition for infectious growth is then $dI/dt>0$ or,

\begin{align}
\begin{split}
+\beta I S /N -\gamma I &> 0 \\
\frac{\beta}{\gamma}&>\frac{N}{S} 
\end{split}
\end{align}

At the beginning of the simulation, the condition for infectious growth becomes
\begin{eqnarray}
{\cal R} \equiv \frac{\beta S_o}{\gamma N}&>&1
\label{eq:rep1}
\end{eqnarray}
which simplifies to ${\cal R} \sim \beta/\gamma$ for large initial susceptible population, $S_o$.  

The same can be done for the SIRI model with little modification \cite{DBLP:journals/corr/CannarellaS14}.

\begin{align}
\begin{split}
\frac{dS}{dt} &= -\beta I S /N \\
\frac{dI}{dt} &= +\beta I S /N -\nu I R/N\\
\frac{dR}{dt} &= +\nu I R
\end{split}
\end{align}
The condition for infectious growth is then $dI/dt>0$ or,
\begin{align}
\begin{split}
+\beta I S /N -\nu I R/N &> 0 \\
\frac{\beta S}{\nu}&>R 
\end{split}
\end{align}
At the start of the simulation, $R=1$, so we have a modified reproduction number and condition for infectious growth,
\begin{eqnarray}
{\cal R} &\equiv& \frac{\beta S_o}{\nu}
\label{eq:rep2}
\end{eqnarray}

\subsection{Statistical Model}
\label{sec:statmodel}

A statistical model of the infectious parameters, $\beta$ and $\gamma$ (or $\nu$), as well as the initial values for the infectious ($I_o$) and susceptible ($S_o$) populations, is attached to the dynamical SIR model.  Uniform prior probabilities are assumed for the parameters and a Python implementation of the affine invariant MCMC ensemble sampler \cite{goodman10} is used to obtain posterior probabilities for the parameter values as well as the correlation between parameters \cite{foreman-mackey13,gelman2004bayesian}.  The MCMC simulation is run for approximately 500000 iterations with a burn-in of 50\%. Following the MCMC iterations, histograms of the parameter traces are used to estimate the posterior probabilities for the parameter values, the uncertainty in those values, and the correlation between parameters in the models.  Calculation of the basic reproduction number \cite{Holme2015}, defined in Equations~\ref{eq:rep1} and ~\ref{eq:rep2}, follows from the estimates of the individual parameters, $\beta$ and $\gamma$ (or $\nu$).  ${\cal R}>1$ denotes an infectious process.

\section{Results}
\label{S:3}

The results fall into three main categories: the dynamics of individual time-series, the collective pattern of the values of $\beta$ and $\gamma$ (or $\nu$ for the SIRI model), and the distribution of the basic reproductive number, ${\cal R}$.

\subsection{Individual traces}

\begin{figure}
\centering
\includegraphics[width=4in]{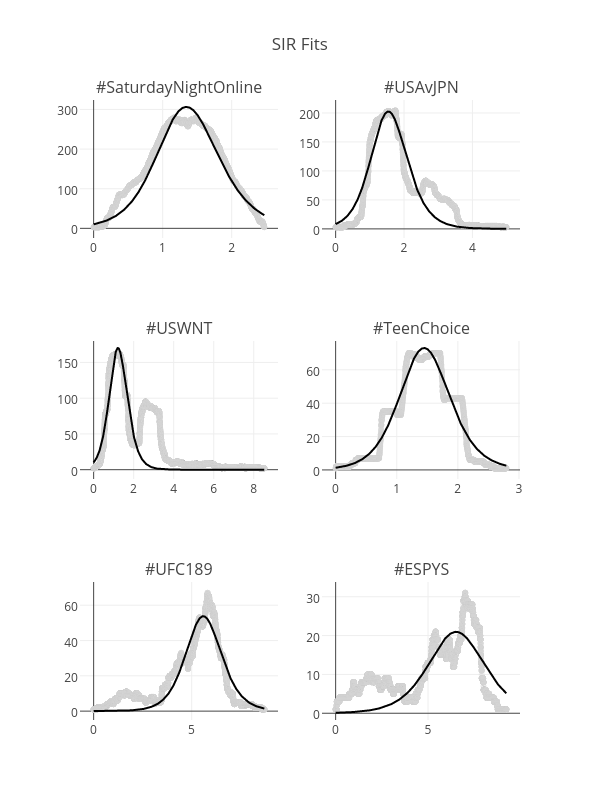}
\caption{Individual traces for selected Twitter hashtags.  Shown are the hashtag counts (gray), smoothed over a one hour time period, and the best fit SIR model (black) for 6 different Twitter hashtags.}
\label{fig:indivtraces}
\end{figure}

Shown in Figure~\ref{fig:indivtraces}, we have a number of sample individual traces of the dynamical model.  For each plot, we show the original data (smoothed over a one-hour time period) and the best-fit dynamical curve drawn from the posterior distribution discovered through the MCMC analysis.  We can immediately see that the SIR model is good at summarizing many of the series, and the dynamics make an approximate sense even in those cases where the data is not smooth (e.g., \texttt{\#TeenChoice}).  In multi-peak cases, the model typically picks out either the first peak (e.g., \texttt{\#USWNT}) or smooths the entire dynamics over a few peaks (e.g., \texttt{\#ESPYS}).

\subsection{Pattern of Parameters}

\begin{figure}[h]
\centering
\begin{tabular}{lr}
{\Large (A)}&\\
&\includegraphics[width=0.5\linewidth]{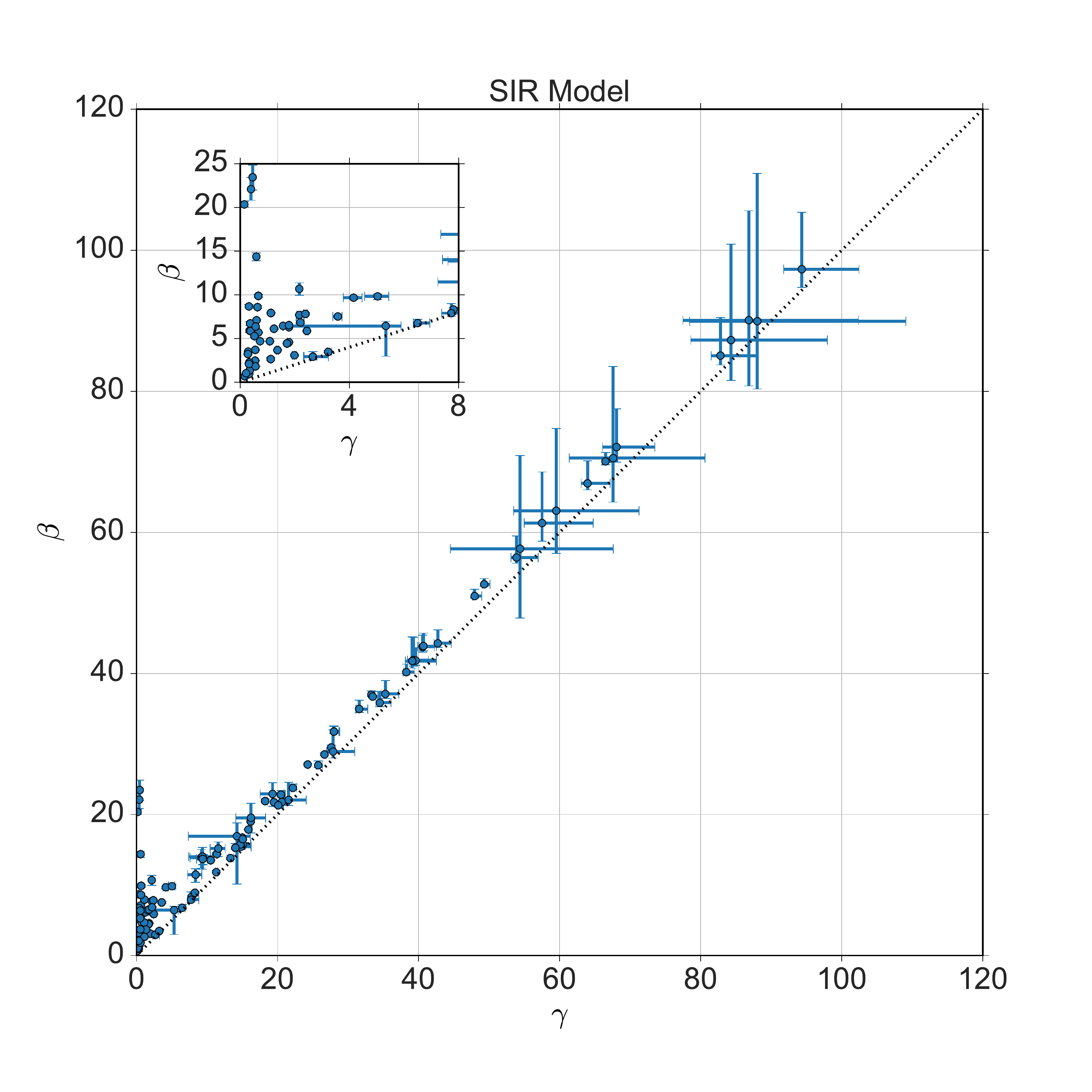}\\
{\Large (B)}&\\
&\includegraphics[width=.5\linewidth]{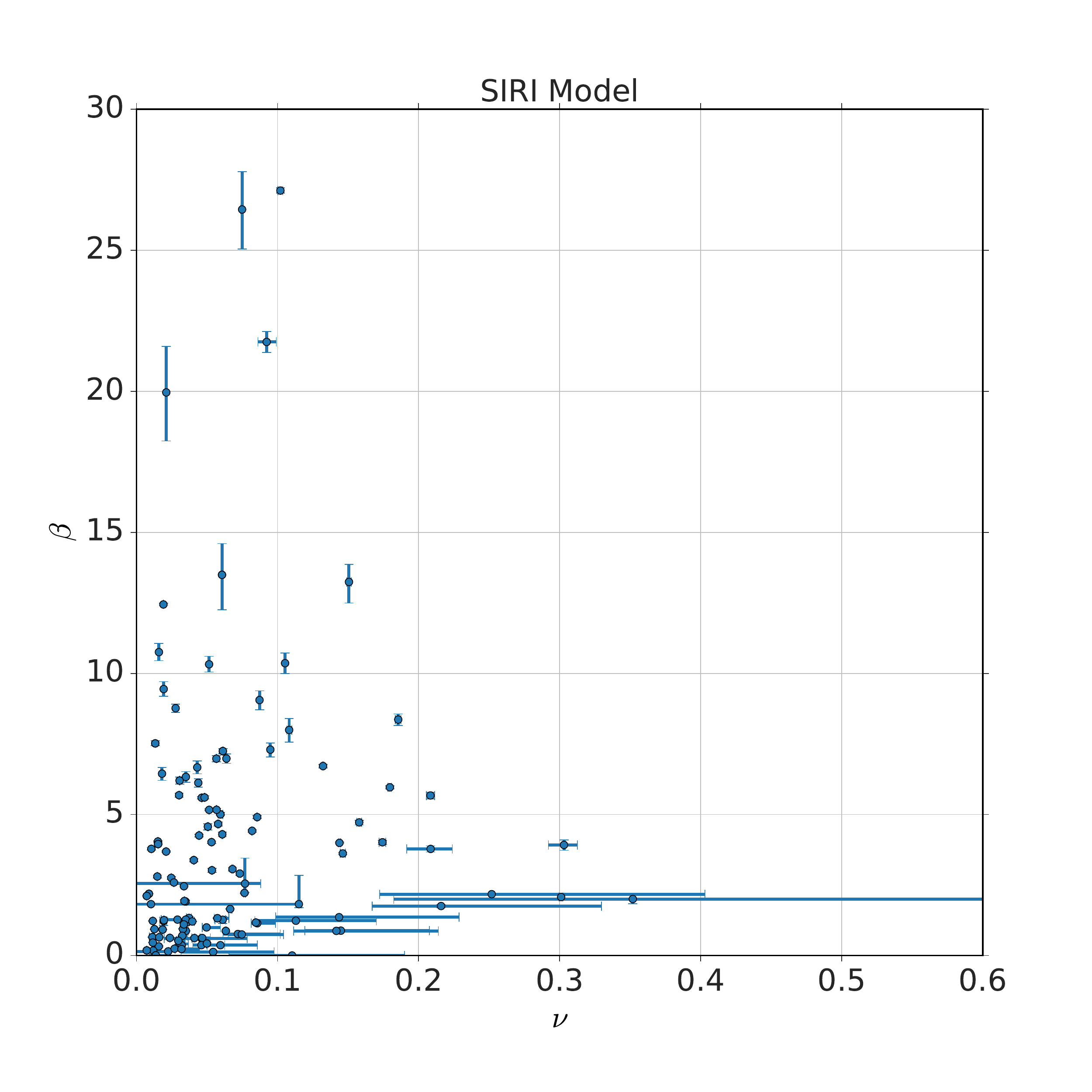}
\end{tabular}
\caption{Plot of $\beta$ vs. $\gamma$ for the SIR model (A) and $\beta$ vs. $\nu$ for the SIRI model (B). All observations lie above the 45$^\circ$ line and are thus represent epidemics to some extent.  The inset on the SIR model expands the lower left-hand corner, where extremely infectious hashtags exist.  The SIRI model (B) consistently results in more infectious hashtags.}
\label{fig:beta-gamma}
\end{figure}

Because hundreds of hashtags were analyzed, we visualize the resulting fits by plotting $\beta$ vs $\gamma$.  Since the basic reproduction number for the SIR model is approximately the ratio of $\beta$ to $\gamma$, as shown in Equation~\ref{eq:rep1}, and ${\cal R}>1$ denotes an infectious disease, then a plot of the points ($\beta$,$\gamma$) for each hashtag should result in a scatter where the infectious topics fall above the 45$^\circ$ line.  The farther away from that line, the more infectious the topic is.  For our analysis of the dynamics of the hashtags collected, the result is shown in Figure~\ref{fig:beta-gamma}.  Two observations are immediately clear.  First, the scatter falls primarily along a single line, implying a very consistent reproduction number for many of the hashtags, slightly above 1.  Second, there is a subpopulation (in the lower left, expanded in the inset) falling well above the 45$^\circ$ line, and are thus quite infectious.  This represents about 20\% of all the tweets analyzed.

\FloatBarrier
\subsection{Reproduction Number}

\begin{figure}
\centering
\begin{tabular}{lr}
{\Large (A)}&\\
\includegraphics[width=.75\linewidth]{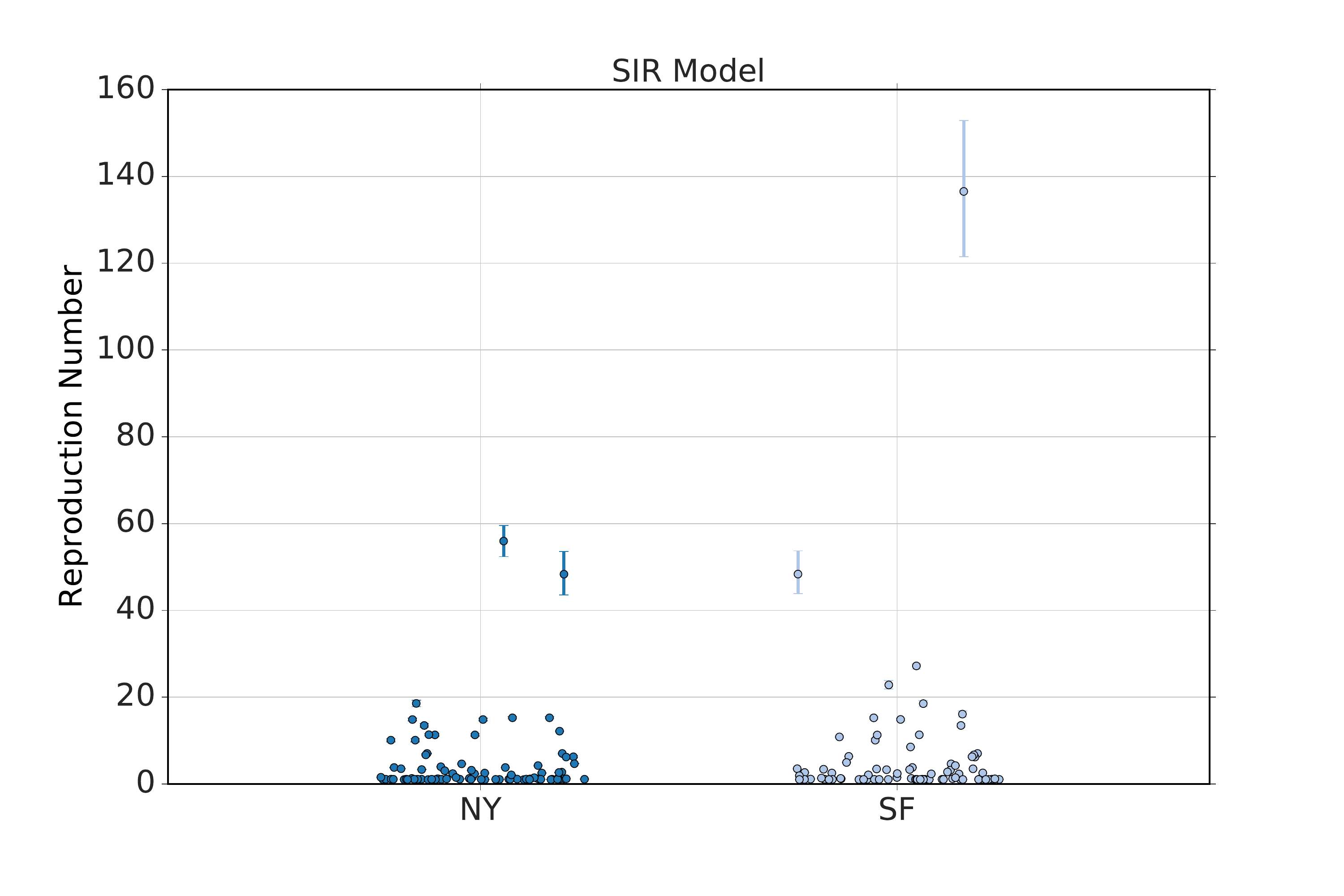} \\
{\Large (B)}&\\
\includegraphics[width=.75\linewidth]{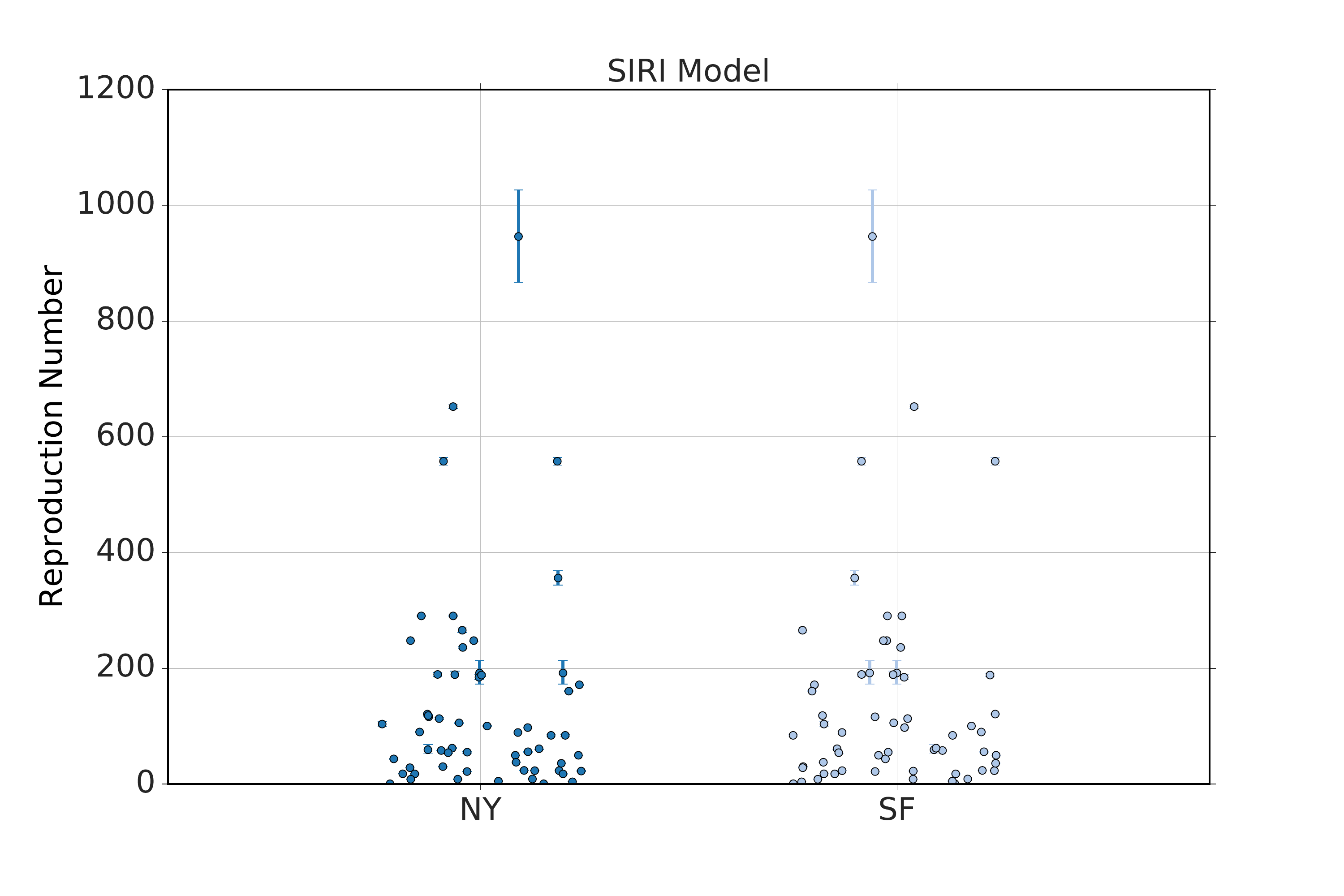}
\end{tabular}
\caption{Distribution of ${\cal R}$ for different geo-locations and dynamical models.  Shown are the values of the reproduction number for the SIR model (A) and the SIRI model (B) for New York tweets and San Francisco tweets.  There is no qualitative difference between the two locations.}
 \label{fig:reproductionnumber}
\end{figure}

Shown in Figure~\ref{fig:reproductionnumber} is the histogram of measured reproduction numbers across all of the trending hashtags.  The range of observed numbers is large, with values consistent with other studies of Twitter hashtag dynamics \cite{eager15}.  There are many with reproduction numbers falling just above ${\cal R}\sim 1$, representing only slightly infectious ideas.  The SIRI model, with the added feedback in the decay of the process, yields parameters with very high infectious rates for the same data as the SIR model.  

\section{Discussion}
\label{S:4}

The results presented here are the first \textit{systematic} look at hashtag dynamics in a Twitter dataset.  Here we demonstrate that the Twitter hashtags fall into two groups: slightly infectious (from an SIR model perspective) with a subset of very infectious topics.  Depending on model assumptions, such as a feedback with ``recovered'' individuals affecting the decay of the topic interest (i.e. the ``infected'' individuals), the best fit can lead to very high infectious conditions.  No significant differences were observed in these conclusions depending on the geography of the hashtags---coming from New York or San Francisco.

Further work on this area is needed to explore the details of these results.  One interesting study would look at the global distribution of the best-fit parameter values, to explore the differences and similarity of topic infectiousness across the globe.  Another would explore the sensitivity of this approach to model assumptions, and to perform a full model comparison in order to confirm the nature of the interactions between the subpopulations of the model.  In the long run, it may be possible to use this approach in order to improve the labeling of ``trending'' topics.





\newpage


\bibliographystyle{model1-num-names}
\bibliography{sources.bib}







\end{document}